\documentstyle[aps,prl]{revtex}
\draft

\newcommand{\gsim}{\lower.7ex\hbox{$\;\stackrel{\textstyle>}{\sim}\;$}}
\newcommand{\lsim}{\lower.7ex\hbox{$\;\stackrel{\textstyle<}{\sim}\;$}}
\def\t{\tilde }
\def\fiv{${\bf 5}+{\bf\overline5}$}
\def\ten{${\bf 10}+{\bf\overline{10}}$}
\def\npb#1 #2 #3 #4 {Nucl.~Phys. {\bf B#1}, #2 (#3)#4 }
\def\plb#1 #2 #3 #4 {Phys.~Lett. {\bf B#1}, #2 (#3)#4 }
\def\prd#1 #2 #3 #4 {Phys.~Rev.  {\bf D#1}, #2 (#3)#4 }
\def\prl#1 #2 #3 #4 {Phys.~Rev.~Lett. {\bf #1}, #2 (#3)#4 }

\begin{document}

\twocolumn[\hsize\textwidth\columnwidth\hsize\csname
@twocolumnfalse\endcsname

\title{New Limits on the SUSY Higgs Boson Mass}

\author{Konstantin T. Matchev}
\address{Theory Group, Fermi National Laboratory, Batavia, IL 60510}

\author{Damien M. Pierce} 
\address{Stanford Linear Accelerator Center, Stanford University,
Stanford, CA 94309}

\maketitle

\begin{abstract}
We present new upper limits on the light Higgs boson mass $m_h$ in
supersymmetric models. We consider two gravity-mediated models (with
and without universal scalar masses) and two gauge-mediated models
(with a \fiv\ or \ten\ messenger sector).  We impose standard
phenomenological constraints, as well as SU(5) Yukawa coupling
unification.  Requiring that the bottom and tau Yukawa couplings meet
at the unification scale to within 15\%, we find the upper limit
$m_h<114$ GeV in the universal supergravity model.  This reverts to
the usual upper bound of 125 GeV with a particular nonuniversality in
the scalar spectrum.  In the \fiv\ gauge-mediated model we find
$m_h<97$ GeV for small $\tan\beta$ and $m_h\simeq 116$ GeV for large
$\tan\beta$, and in the \ten\ model we find $m_h<94$ GeV. We discuss
the implications for upcoming searches at LEP-II and the Tevatron.
\end{abstract}
\pacs{PACS numbers: 11.30.Pb, 12.10.Kt, 14.80.Cp
\hfill SLAC-PUB-7821~~~FERMILAB-PUB-98/143-T}

]

\setcounter{footnote}{0}
\setcounter{page}{1}

If weak-scale supersymmetry (SUSY) exists it may be a challenge to
discover.  The superpartners may all be so heavy that they do not
appreciably affect any low energy observables, and are below threshold
for production at LEP-II and the Tevatron. In that case they will go
undiscovered until the LHC turns on in 2005. However, one of the most
robust and enticing hallmarks of supersymmetric models is the
prediction of a light Higgs boson. At tree level, $m_h\leq M_Z$, but
it receives large radiative corrections from top and stop loops
\cite{1 loop Higgs}. Naturalness suggests that the (third generation)
squarks should not be too heavy. If we impose that the (top) squark
masses are below 1.2 TeV, the upper limit on $m_h$ is about 125 GeV
(including recent two-loop corrections \cite{2 loop Higgs}). The
largest possible value is obtained with heavy stops and large squark
mixing.

Supersymmetry goes hand in hand with grand unification. The couple of
percent discrepancy in gauge coupling unification finds a ready
explanation in grand unified theory (GUT) models, from GUT threshold
effects. In typical GUT models the bottom and tau Yukawa couplings
($\lambda_b$ and $\lambda_\tau$) are predicted to unify as well.  At
leading order this only happens for either very small ($\lsim 2$) or
rather large ($\sim \!m_t/m_b$) values of $\tan\beta$ ($\tan\beta$ is
the ratio of expectation values of the two Higgs doublets).  

We take the muon decay constant, the $Z$-boson mass, the fermion
masses, $\tan\beta$, and the strong and electromagnetic couplings as
inputs to determine $\lambda_b$ and $\lambda_\tau$ at the GUT scale
(the scale where the U(1) and SU(2) gauge couplings meet). We define
the Yukawa coupling mismatch at the GUT scale to be $\varepsilon_b
\equiv ( \lambda_b - \lambda_\tau )/ \lambda_\tau$, and, allowing for
variations in the input parameters and GUT scale threshold
corrections, conservatively expect $\varepsilon_b$ to be less than
15\% in magnitude.

At next-to-leading order bottom-tau unification is sensitive to the
supersymmetric spectrum through radiative corrections. The corrections
to $\lambda_b$ are enhanced at large $\tan\beta$ and can be quite
large \cite{COPW,Hall,BMPZ_NPB}. They broaden the region at large
$\tan\beta$ where exact unification is possible to
15$\lsim\tan\beta\lsim$50.  The branching ratio ${\rm B}(B\rightarrow
X_s\gamma)$ also receives large $\tan\beta$ enhanced corrections. The
requirements of Yukawa unification and compliance with the
B$(B\rightarrow X_s\gamma)$ measurement tend to conflict with each
other. Depending on the model, imposing both constraints can single
out a very particular parameter space, resulting in predictions for
the superpartner and Higgs boson masses. In this letter we examine the
Higgs boson mass predictions in four supersymmetric models -- two
gravity-mediated models (with and without universal scalar masses),
and two gauge-mediated models (with a \fiv\ or \ten\ messenger
sector).  Yukawa coupling unification together with the $b\rightarrow
s\gamma$ constraint has been previously discussed within the context
of the gravity-mediated models in Ref.~\cite{COPW}, but no conclusions
about the light Higgs boson mass were drawn. Ref.~\cite{Chankowski}
uses fine-tuning criteria in addition to the ${\rm B}(B\rightarrow
X_s\gamma)$ constraint to derive some limits on the light Higgs boson
mass.

In each model we randomly pick points in the supersymmetric parameter
space (the parameter spaces are discussed below). At each point the
$Z$-boson mass, the top-quark mass, and the electromagnetic and strong
couplings at the $Z$-scale are determined in a global fit to precision
data. We construct a $\chi^2$ function and minimize it with respect to
the four standard model inputs. The $\chi^2$ function contains 30
electroweak precision observables, and B$(B\rightarrow
X_s\gamma)$. The list of observables, the measurements we use, and
further details are given in Ref.~\cite{EP}. We set $m_b({\rm
pole})=4.9$ GeV.

We impose a number of phenomenological constraints at each point in
parameter space. We require radiative electroweak symmetry breaking
and determine the CP-odd Higgs boson mass $m_A$ and the absolute
value of the Higgsino mass parameter $\mu$ to full one-loop order
\cite{BMPZ_NPB}. Very large values of $\tan\beta$ are excluded by this
constraint. We require that all Yukawa couplings remain perturbative
up to the GUT scale. This rules out very small values of
$\tan\beta$. Finally, we require that all the superpartner and Higgs
boson masses are above the bounds set by direct particle searches.

We calculate the gauge and Yukawa couplings using the full one-loop
threshold corrections~\cite{BMPZ_NPB} and two-loop renormalization
group equations \cite{2 loop RGEs}.  The parameter dependence of the
$\lambda_b$ corrections can be understood from the simplified
approximation
\begin{equation}
{\delta \lambda_b \over \lambda_b } \ \simeq\
- {1\over 16\pi^2} \biggl({8\over3}g_3^2 m_{\t g} + \lambda_t^2 A_t\biggr)
{\mu \tan\beta \over m_{\t q}^2}~.
\label{delta_yb}
\end{equation}
The first (second) term is the gluino-sbottom (chargino-stop) loop
contribution. $m_{\t q}$ is an average (stop or sbottom) squark mass,
$m_{\t g}$ is the gluino mass, $A_t$ is the stop-stop-Higgs trilinear
coupling and $g_3$ ($\lambda_t$) is the strong (top Yukawa)
coupling. In a leading-order analysis, where the corrections
(\ref{delta_yb}) are neglected, $\lambda_b$ and $\lambda_\tau$ unify
well below the GUT scale for intermediate values of $\tan\beta$.  With
$\mu>0$ the corrections (\ref{delta_yb}) make this situation worse, so
that with $\tan\beta>2$ $\varepsilon_b$ falls in the range $-20$ to
$-60\%$.  This discrepancy is larger than can be accounted for in
realistic GUT models \cite{Wright,BMPZ_PRL}.  Also, variations in the
input parameters $\Delta m_t=\pm3$ GeV, $\Delta m_b=\pm0.15$ GeV and
$\Delta \alpha_s=\pm0.003$ result in $\Delta\varepsilon_b=\pm1\%$,
$\pm3\%$, and $\mp3\%$, respectively. With $\mu<0$ the threshold
corrections (\ref{delta_yb}) help Yukawa unification by increasing
$\lambda_b$ at the weak scale, thus delaying its unification with
$\lambda_\tau$ to higher scales. Our conservative requirement
$|\varepsilon_b|<15\%$ restricts us to either $\tan\beta<2$ with $\mu$
of either sign, or $\tan\beta\gsim5$ and $\mu<0$.

The allowed values of $A_t$ play a central role in our
discussion. Each model allows for a different range of values of
$A_t$, with corresponding implications. We start by discussing the
results in the gravity-mediated model with universal soft parameters
(mSUGRA). In this model three inputs are specified at the GUT scale.
They are a universal scalar mass $M_0$, a universal gaugino mass
$M_{1/2}$, and a universal trilinear scalar coupling $A_0$. The
remaining two inputs are $\tan\beta$ and the sign of $\mu$.

\begin{figure}[tb]
\vbox{\kern2.4in\includegraphics{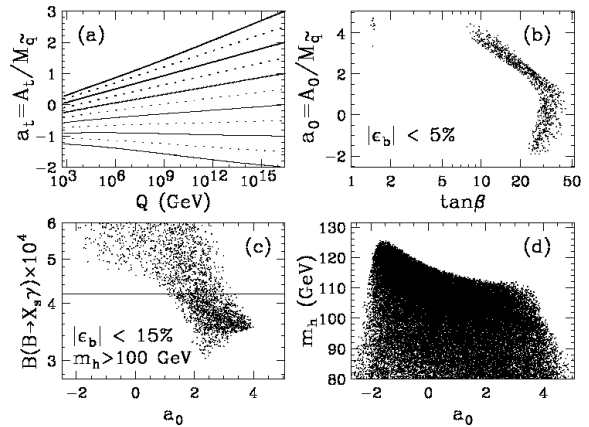}}
\caption{\small (a) Renormalization group trajectories of $a_t$ for
$M_0=500$ GeV, $M_{1/2}=200$ GeV, $\tan\beta=20$ and $\mu<0$; (b) The
stop mixing parameter $a_0$ vs. $\tan\beta$ with $|\varepsilon_b| <
5\%$; (c) The branching ratio ${\rm B}(B\rightarrow X_s\gamma)$ versus
$a_0$, with the additional constraints $|\varepsilon_b|<15\%$ and
$m_h>100$ GeV.  (d) The light Higgs boson mass as a function of $a_0$,
scanned over the mSUGRA parameter space with no additional
constraints.}
\label{fig1}
\end{figure}

Because of the large top Yukawa coupling, the value of $A_t$ at the
weak scale exhibits a quasi-fixed point behavior. Hence, the
sensitivity to its high scale boundary condition is reduced.  The
quasi-fixed point behavior is illustrated in Fig.~\ref{fig1}(a), where
we plot the dimensionless parameter $a_t \equiv A_t/M_{\tilde q}$ as a
function of the renormalization scale $Q$ in the mSUGRA model
($M_{\tilde q}\equiv\sqrt{M_0^2+4M_{1/2}^2}$ is approximately equal to
the first or second generation squark mass). We see that $A_t$ tends
to be negative at the weak scale. In that case the stop-chargino
contribution in Eq.~(\ref{delta_yb}) partially cancels the
sbottom-gluino contribution.  At intermediate values of $\tan\beta$
($10\lsim\tan\beta\lsim20$) Yukawa unification requires that the
correction (\ref{delta_yb}) be maximized. This happens when $A_t>0$,
so that the stop-chargino contribution adds constructively to the
sbottom-gluino contribution. Hence, at intermediate $\tan\beta$ large
and positive values of $a_0\equiv A_0/M_{\tilde q}$ are necessary in
the mSUGRA model. This is illustrated in Fig.~\ref{fig1}(b), where we
show the results of a scan over the mSUGRA parameter space with the
requirement that $|\varepsilon_b|<5\%$. The figure shows a striking
correlation between $a_0$ and $\tan\beta$ at intermediate values of
$\tan\beta$. At large $\tan\beta$ the $\tan\beta$ enhancement in
Eq.~(\ref{delta_yb}) is by itself enough for successful Yukawa
unification. In fact, at some points cancellation between the two
terms in (\ref{delta_yb}) is necessary, so small or negative values of
$a_0$ are preferred. We also see from Fig.~\ref{fig1}(b) that in the
small $\tan\beta$ region large positive $A_0$ is required, signifying
that the corrections in Eq.~(\ref{delta_yb}) are relevant.  The points
in this region have values of the Higgs boson mass below 100 GeV.

The CLEO collaboration's 95\% upper bound on the $b\rightarrow
s\gamma$ rate, B$(B\rightarrow X_s\gamma)<4.2\times10^{-4}$
\cite{CLEO}, imposes
strong constraints on supersymmetric models. If $\mu<0$ the
chargino-stop and Higgs boson contributions to the $b\rightarrow
s\gamma$ amplitude add constructively to the SM amplitude. Due to the
$\tan\beta$ enhancement of the chargino loop contribution, very large
total amplitudes can result, leading to predictions for ${\rm
B}(B\rightarrow X_s\gamma)$ well above the upper bound. As a result,
significant regions of the SUSY parameter space are excluded. We can
identify those by considering the following approximate formula for
the leading supersymmetric corrections to the $O_7$ operator
coefficient. With $\mu<0$ and $\tan\beta$ large, we have
\begin{equation}
\delta C_7(M_W) \simeq -{3\tan\beta\over16\pi^2|\mu|}\biggl[{M_W^2\over
M_2}-{1\over2}{A_tm_t^2\over m_{\tilde t}^2}\biggr]~,
\label{C7}
\end{equation}
where the first (second) term is the contribution from the $\t t_L-\t
\chi^+$ ($\t t-\t h^+$) loop (we work to first order in stop and
chargino mixing). $M_W$ ($m_t$) is the $W$-boson (top-quark) mass,
$M_2$ is the SU(2) gaugino soft mass, and $m_{\tilde t}$ is the
average stop mass.  If $A_t>0$ there is destructive interference
between the two terms, and the supersymmetric contribution to the
$b\rightarrow s\gamma$ amplitude is reduced.  In Fig.~\ref{fig1}(c) we
show the full one-loop prediction for B$(B\rightarrow X_s\gamma)$ in
the mSUGRA model, subject to the $b-\tau$ unification constraint
$|\varepsilon_b|<15\%$. As expected, the rate is suppressed for large
and positive values of $a_0$.

We see from Fig.~\ref{fig1}(c) that in the mSUGRA model with
$\tan\beta>2$, reasonably small $|\varepsilon_b|$ and ${\rm
B}(B\rightarrow X_s\gamma)$ can occur only for relatively large and
positive $a_0$ ($a_0>1.1$).  Because of the focusing towards negative
values, the resulting values of $a_t$ at the squark mass scale are
rather small ($-0.4<a_t<0.7$).  Hence, top squark mixing is suppressed
and the corrections to the Higgs boson mass are minimized.  The
scatter plot in Fig.~\ref{fig1}(d) shows $m_h$ vs. $a_0$ in the mSUGRA
model. The Higgs boson mass is maximal at $a_0=-1.7$ and decreases
with increasing $a_0$.

\begin{figure}[tb]
\vbox{\kern2.4in\includegraphics{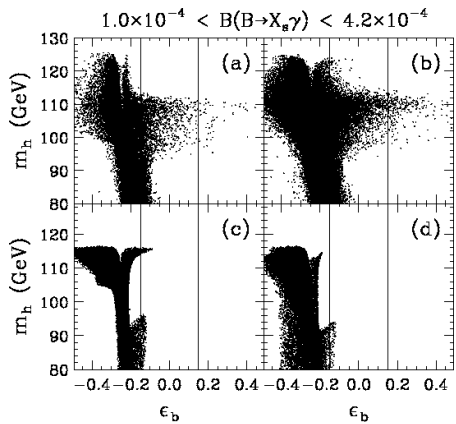}}
\caption{\small Scatter plots of $m_h$ vs.  $\varepsilon_b$ in four
supersymmetric models: SUGRA with (a) universal or (b) nonuniversal
scalar masses; and minimal gauge mediation with a (c) \fiv\ or (d)
\ten\ messenger sector. In each case, we require
$1.0\times10^{-4}<{\rm B}(B\rightarrow X_s\gamma)<4.2\times10^{-4}$.
The vertical lines on the plots delineate the region
$|\varepsilon_b|<15\%$.}
\label{fig2}
\end{figure}

In Fig.~\ref{fig2}(a) we show the scatter plot of $m_h$ vs.
$\varepsilon_b$ in the mSUGRA model. We have imposed the $b\rightarrow
s\gamma$ constraint in Fig.~2. The vertical lines indicate the region
$|\varepsilon_b|<15\%$. We see that the Higgs boson mass is below 114
GeV in this region.

The mSUGRA model suffers from the rather ad hoc assumption of scalar
mass unification. While we see no compelling justification for this
boundary condition, if it did apply it would naturally hold at the
Planck scale. The effects of running between the Planck scale and the
GUT scale can be significant \cite{P&P}. Regardless of the boundary
condition at the Planck scale, a GUT symmetry will ensure that
GUT multiplets remain degenerate above the GUT scale. In SU(5), the
parameter space at the GUT scale includes the soft mass parameters
$M_{H_1}$, $M_{H_2}$, $M_5$ and $M_{10}$, corresponding to the ${\bf
\overline 5}$ and {\bf 5} representations of Higgs fields, and the
${\bf \overline 5}$ and {\bf 10} representations of sfermion fields,
respectively. (Although we impose generation independence of the $M_5$
and $M_{10}$ masses, the phenomenology we consider here only depends
on the third generation scalar masses.)

Because of the larger parameter space of the nonuniversal SU(5) model,
the upper limit on the Higgs boson mass, including the ${\rm
B}(B\rightarrow X_s\gamma)$ and approximate bottom-tau unification
constraints, reverts to the general upper limit of 125 GeV (see
Fig.~\ref{fig2}(b)). We can contrast the situation here with the
standard mSUGRA case. Large splitting between $M_{H_2}$ and $M_{10}$
can lead to much larger values of $\mu$ and $m_A$ for a given
$\tan\beta$. The larger value of $\mu$ gives larger corrections to the
bottom Yukawa coupling, making bottom-tau unification possible for
smaller values of $\tan\beta$. The smaller values of $\tan\beta$, with
larger $\mu$ and $m_A$, lead to reductions in the supersymmetric
contributions to the $b\rightarrow s\gamma$ amplitude. This, in turn,
allows compliance with the B$(B\rightarrow X_s\gamma)$ upper bound
with large and negative values of $A_0$. Such $A_0$ values result in
large squark mixing contributions to the Higgs boson mass (see
Fig.~1(d)). The points with the largest $m_h$ have $M_{10}\simeq1$
TeV, $M_{1/2}\simeq 400\pm100$ GeV, $\tan\beta\simeq 14\pm3$, $A_0$ in
the range $-1$ to $-2$ TeV and, typically, $M_{H_2}\lsim 100$ GeV.

Gauge mediation is an attractive alternative to gravity-mediated
supersymmetry breaking. One of the nice features of gauge-mediated
models is the automatic scalar mass degeneracy. All sfermions with
identical quantum numbers have the same mass at the messenger
scale. This provides a natural solution to the supersymmetric flavor
problem.

In order to preserve gauge coupling unification we consider two models
with full SU(5) messenger sector representations, the \fiv\ and \ten\
models. We assume a minimal Higgs sector, where the mechanism which
gives rise to the $B$ and $\mu$ terms does not give additional
contributions to the scalar masses. In the canonical models \cite{DNS}
the interactions between the dynamical supersymmetry breaking sector
and a standard model singlet give rise to a vev in its scalar and $F$
components. The coupling of the singlet to the messenger fields
results in supersymmetry breaking and conserving messenger masses. To
determine the effective theory below the messenger mass scale, the
messenger fields are integrated out. The MSSM superpartners then
receive masses proportional to $\Lambda=F/S$, where $F$ ($S$) is the
singlet $F$-term (scalar) vev. At this order, there is no $A$-term
generated. Hence, we set $A_0=0$ at the messenger scale. The messenger
scale determines the amount of running of the soft parameters. The
smallest allowed value of the messenger scale is $\Lambda$. Since we
do not want the gravity-mediated contributions to the scalar masses to
spoil the solution to the supersymmetric flavor problem, we suppress
the gravity-mediated contributions by requiring the messenger scale to
be below $M_{\rm GUT}/10$.

In the mSUGRA model we found that the B($B\rightarrow X_s\gamma$) and
bottom-tau unification constraints required $a_0>1.1$ at intermediate
to large $\tan\beta$. Since the gauge-mediated models have $a_0=0$,
one would expect that these models would not be compatible with the
constraints at intermediate to large $\tan\beta$ if the spectrum did
not significantly differ from the mSUGRA model. However, it is well
known that the spectra in gauge- and gravity-mediated models can be
quite different \cite{BMPZ_PRD}. For example, in the \fiv\
gauge-mediated model, the scalar masses and the $\mu$-term are
significantly heavier for a given gaugino mass than in the mSUGRA
model. Just as in the nonuniversal model, the larger $\mu$ allows for
bottom-tau unification with smaller values of $\tan\beta$, and the
reduced $\tan\beta$ and larger $\mu$ and $m_A$ suppress the
supersymmetric contribution to the $b\rightarrow s\gamma$
amplitude. Hence, in the \fiv\ model there is a small amount of
parameter space at intermediate $\tan\beta$ where the constraints are
satisfied, even though $A_t<0$. In this region we find the prediction
$m_h\simeq116$ GeV.  In the small $\tan\beta$ region $m_h<97$ GeV.
The results are shown in Fig.~2(c).

For a given gaugino mass, larger messenger sector representations
result in lighter scalar masses. Hence, the \ten\ model has relatively
lighter scalars than the \fiv\ model. The lighter scalars make Yukawa
unification more difficult, and readily result in too large values of
B$(B\rightarrow X_s\gamma)$ at intermediate to large $\tan\beta$. As
can be seen in Fig.~2(d), the two constraints taken together exclude
the \ten\ model outright for intermediate to large values of
$\tan\beta$. The only allowed points with $|\varepsilon_b|<15\%$
correspond to values of $\tan\beta<2$. In this region $m_h$ is less
than 94 GeV.

Our results are particularly interesting in light of the upcoming
Higgs boson searches at LEP and the Tevatron. LEP-II should be able to
either discover or rule out a light Higgs boson up to about 105
GeV. If LEP finds a Higgs boson {\em heavier} than 96 (94) GeV, the
minimal \fiv\ (\ten) gauge-mediated model will require some
modification in order to be compatible with bottom-tau
unification. If, on the other hand, LEP does not find a light Higgs
boson, the Yukawa unification criterion excludes the minimal \ten\
gauge-mediated model.

What is more, if bottom-tau unification is taken seriously, upcoming
runs at the Tevatron stand a chance to explore both the mSUGRA and
minimal \fiv\ gauge-mediated models. The Tevatron reach in $m_h$ as a
function of its total integrated luminosity is currently under active
investigation and no definite conclusions can be made at this point,
but the upper limits of 114 and 116 GeV, correspondingly, can serve as
important benchmarks in the design of an extended Run 2. Finally, if
the Tevatron can place a limit on the Higgs boson mass above 116 GeV,
this would point towards particular nonunified scenarios in the
gravity-mediated models, and exclude Yukawa coupling unification in
minimal gauge-mediation altogether.

\section*{Acknowledgments}
KTM wishes to thank the SLAC theory group for its hospitality during a
recent visit. We thank J. Erler for his contribution to the global fit
program. Work of KTM and DMP supported by Department of Energy
contracts DE-AC02-76CH03000 and DE-AC03-76SF00515, respectively.


\begin{references}

\bibitem{1 loop Higgs} 
H.~Haber, {\tt hep-ph/9707213} and references therein.

\bibitem{2 loop Higgs} 
J.R.~Espinosa and M.~Quiros, \plb 266 389 1991 ;
J.~Kodaira, Y.~Yasui and K.~Sasaki, \prd 50 7035 1994 ;
R.~Hempfling and A.~Hoang, \plb 331 99 1994 ;
J.A.~Casas, J.R.~Espinosa, M.~Quiros and A.~ Riotto, \npb 436 3 1995 ;
M.~Carena, M.~Quiros and C.E.M.~Wagner, \npb 461 407 1996 ;
S.~Heinemeyer, W.~Hollik and G.~Weiglein, {\tt hep-ph/9803277}.

\bibitem{COPW}
M.~Carena, M.~Olechowski, S.~Pokorski and C.E.M.~Wagner, \npb 426 269
1994 ; M.~Carena, and C.E.M.~Wagner, {\tt hep-ph/9407209}.

\bibitem{Hall}
L.~Hall, R.~Rattazzi and U.~Sarid, \prd 50 7048 1994 .

\bibitem{BMPZ_NPB}
J.~Bagger, K.~Matchev, D.~Pierce and R.-J.~Zhang, \npb 491 3 1997 .

\bibitem{Chankowski}
P.~Chankowski and S.~Pokorski, {\tt hep-ph/9702431}.

\bibitem{EP}
J.~Erler and D.~Pierce, {\tt hep-ph/9801238}, to appear in
Nucl. Phys. {\bf B}.

\bibitem{2 loop RGEs}
Y. Yamada, \prd 50 3537 1994 ;
S. Martin and M. Vaughn, \plb 318 331 1993 ;
{\it ibid.}, \prd 50 2282 1994 ;
I. Jack and D.R.T. Jones, \plb 333 372 1994 .

\bibitem{Wright}
B.D.~Wright, {\tt hep-ph/9404217}.

\bibitem{BMPZ_PRL}
J.~Bagger, K.~Matchev, D.~Pierce and R.-J.~Zhang, \prl 78 2497 1997 .

\bibitem{CLEO}
CLEO collaboration: M.S.~Alam {\it et al.}, \prl 74 2885 1995 .

\bibitem{P&P}
N.~Polonsky and A.~Pomarol, \prl 73 2292 1994 ,\prd 51 6532 1995 .

\bibitem{DNS}
M.~Dine and A.~Nelson, \prd 48 1277 1993 ;
M.~Dine, A.~Nelson and Y.~Shirman, \prd 51 1362 1995 ;
M.~Dine, A.~Nelson, Y.~Nir and Y.~Shirman, \prd 53 2658 1996 .

\bibitem{BMPZ_PRD}
J.~Bagger, K.~Matchev, D.~Pierce and R.-J.~Zhang, \prd 55 437 1997 .

\end{references}
\end{document}